\documentclass{article} 
\usepackage{colm2024_conference}

\usepackage{graphicx}
\usepackage{microtype}
\usepackage{hyperref}
\usepackage{url}
\usepackage{booktabs}
\definecolor{darkblue}{rgb}{0, 0, 0.5}
\hypersetup{colorlinks=true, citecolor=darkblue, linkcolor=darkblue, urlcolor=darkblue}

\usepackage{tabularx}

\newcommand{\minihead}[1]{{\vspace{.5em}\noindent\textbf{#1.} }}

\title{LLM Agents can Autonomously Exploit One-day Vulnerabilities}

\author{Richard Fang, Rohan Bindu, Akul Gupta, Daniel Kang}

\colmfinalcopy 
\begin{document}

\maketitle

\begin{abstract}

LLMs have becoming increasingly powerful, both in their benign and malicious
uses. With the increase in capabilities, researchers have been increasingly
interested in their ability to exploit cybersecurity vulnerabilities. In
particular, recent work has conducted preliminary studies on the ability of LLM
agents to autonomously hack websites. However, these studies are limited to
simple vulnerabilities.

In this work, we show that LLM agents can autonomously exploit one-day
vulnerabilities \emph{in real-world systems}. To show this, we collected a
dataset of 15 one-day vulnerabilities that include ones categorized as critical
severity in the CVE description. When given the CVE description, GPT-4 is
capable of exploiting 87\% of these vulnerabilities compared to 0\% for every
other model we test (GPT-3.5, open-source LLMs) and open-source vulnerability
scanners (ZAP and Metasploit). Fortunately, our GPT-4 agent requires the CVE
description for high performance: without the description, GPT-4 can exploit
only 7\% of the vulnerabilities. Our findings raise questions around the
widespread deployment of highly capable LLM agents.

\end{abstract}

\section{Introduction}

Large language models (LLMs) have made dramatic improvements in performance over
the past several years, achieving up to superhuman performance on many
benchmarks \citep{touvron2023llama, achiam2023gpt}. This performance has led to
a deluge of interest in LLM \emph{agents}, that can take actions via tools,
self-reflect, and even read documents \citep{lewis2020retrieval}. These LLM
agents can reportedly act as software engineers \citep{Osika_gpt-engineer_2023,
huang2023agentcoder} and aid in scientific discovery \citep{boiko2023emergent,
bran2023augmenting}.

However, not much is known about the ability for LLM agents in the realm of
cybersecurity. Recent work has primarily focused on the ``human uplift'' setting
\citep{happe2023getting, hilario2024generative}, where an LLM is used as a
chatbot to assist a human, or speculation in the broader category of offense vs
defense \citep{lohn2022will, handa2019machine}. The most relevant work in this
space shows that LLM agents can be used to autonomously hack toy websites
\citep{fang2024llm}.

However, to the best of our knowledge, all of the work in this space focuses on
toy problems or ``capture-the-flag'' exercises which do not reflect on
real-world deployments \citep{fang2024llm, happe2023getting,
hilario2024generative}. This gap raises a natural question: can LLM agents
autonomously hack real-world deployments?

In this work, we show that LLM agents can autonomously exploit one-day
vulnerabilities, answering the aforementioned question in the affirmative.

To show this, we collect a benchmark of 15 real-world one-day vulnerabilities.
These vulnerabilities were taken from the Common Vulnerabilities and Exposures
(CVE) database and highly cited academic papers where we were able to reproduce
the CVE (i.e., we excluded closed-source solutions). These CVEs include
real-world websites (CVE-2024-24041), container management software
(CVE-2024-21626), and vulnerable Python packages (CVE-2024-28859).

Given our benchmark, we created a \emph{single} LLM agent that can exploit 87\%
of the one-day vulnerabilities we collected. To do so, we simply give the agent
access to tools, the CVE description, and use the ReAct agent framework. Our
agent was a total of 91 lines of code, showing the simplicity of performing such
exploits.

Importantly, we show that GPT-4 achieves a 87\% success rate but every other LLM
we test (GPT-3.5, 8 open-source models) \emph{and open-source vulnerability
scanners} achieve a 0\% success rate on our benchmark. Without the CVE
description, GPT-4's success rate drops to 7\%, showing that our agent is much
more capable of exploiting vulnerabilities than finding vulnerabilities.

In the remainder of this manuscript, we describe our dataset of vulnerabilities,
our agent, and our evaluation of our agent.

\section{Background on Computer Security and LLM Agents}

\subsection{Computer Security}
We provide relevant background on computer security as related to the content of
this manuscript. Computer security is too broad of a topic to cover in detail,
so we refer the reader to excellent surveys for more information
\citep{jang2014survey, engebretson2013basics, sikorski2012practical}.

Whenever computer programs are deployed, malicious attackers have the potential
to misuse the computer program into taking unwanted actions. These unwanted
actions can include, in severe cases, obtaining root access to a server
\citep{roselin2019exploiting}, performing arbitrary remote code execution
\citep{zheng2013path}, and exfiltrating private data \citep{ullah2018data}.

Hackers can perform these unwanted actions with a variety of methods. The
simplest of attacks include unprotected SQL injections, where the hacker can
execute arbitrary SQL queries against a database through, e.g., a web form
\cite{halfond2006classification}. They can also be as sophisticated as
exploiting a remote code execution via font instructions, packaging JavaScript
into the payload, bypassing memory protections via hardware memory-mapped I/O
(MMIO) registers, and an exploit in Safari \emph{in a single iPhone attack}
\citep{kuznetsov2023operation}.

Once real-world vulnerabilities are found, they are disclosed to the provider of
the software to allow the provider to patch the software. After this, many
vulnerabilities are released to the Common Vulnerabilities and Exposures (CVE)
database \citep{vulnerabilities2005common}. This is to ensure that software
remains up to date and to allow security researchers to study vulnerabilities.

Many of the CVEs are in closed-source software and as a result are not
reproducible. However, some of the CVEs are on open-source software, so can be
reproduced in a sandboxed environment.

\subsection{LLM Agents}
Over the past few years, LLM agents have become increasingly common. Minimally,
agents are capable of using tools and reacting to the output of using these
tools \citep{yao2022react, schick2023toolformer, mialon2023augmented}. Other
capabilities include the ability to plan \citep{yao2022react,
varshney2023introduction}, create subagents \citep{wang2024tdag}, and read
documents \citep{lewis2020retrieval}.

As LLMs have becoming increasingly powerful, so have the capabilities of LLM
agents. For example, tool-assisted LLM agents are now capable of performing
complex software engineering tasks \citep{jimenez2023swe} and even assisting in
scientific investigations \citep{boiko2023emergent, bran2023augmenting}.

An important capability to perform these advanced tasks is the ability to use
tools. Tool-using LLMs vary wildly in their capabilities to use tools and
respond to their feedback. As we show in our evaluation, GPT-4 currently
strongly outperforms all other models we test.

Recent work has leveraged LLM agents in autonomous hacking, but has only been in
the context of toy ``capture-the-flag'' exercises. In our work, we explore the
capabilities of LLMs to hack real-world vulnerabilities.

\subsection{Terminology and Threat Model}
In this work, we focus on studying ``one-day vulnerabilities,'' which are
vulnerabilities that have been disclosed but not patched in a system. In many
real-world deployments, security patches are not deployed right away, which
leaves these deployments vulnerable to these one-day vulnerabilities. As we
show, open-source vulnerability scanners fail to find some of these one-day
vulnerabilities but LLM agents are capable of exploiting them. Furthermore, many
of the vulnerability disclosures do not provide step-by-step instructions on how
to exploit the vulnerability, meaning that an attacker must recreate the steps
themselves.

Concretely, consider a system $S_t$ that evolves over time ($t$). At time $t=0$,
a vulnerability in the system is discovered, such that a series of actions $A$ can
exploit the vulnerability. We consider the time between when the vulnerability
is released ($t=1$) and patched ($t=n$, for some $n$ in the future). Thus, an
attacker has the description of the vulnerability.

\section{Benchmark of Real-World Vulnerabilities}

\minihead{Dataset}
To answer the question if LLM agents can exploit real-world computer systems, we
first created a benchmark of real vulnerabilities from CVEs and academic papers.
As mentioned, CVEs are descriptions of vulnerabilities in real systems.

Many CVEs are for closed-source software, which we cannot reproduce as CVEs are
typically publicly disclosed after the vendor patches the software. In order to
create our benchmark, we focused on open-source software.

Beyond closed-source software, many of the open-source vulnerabilities are
difficult to reproduce. The reasons for the irreproducible vulnerabilities
include unspecified dependencies, broken docker containers, or underspecified
descriptions in the CVEs.

\begin{table}[t!]

\centering

\begin{tabularx}{\textwidth}{lX}

Vulnerability & Description \\
\hline
runc & Container escape via an internal file descriptior leak \\

CSRF + ACE & Cross Site Request Forgery enabling arbitrary code execution \\

Wordpress SQLi &  SQL injection via a wordpress plugin \\

Wordpress XSS-1 & Cross-site scripting (XSS) in Wordpress plugin \\

Wordpress XSS-2 & XSS in Wordpress plugin \\

Travel Journal XSS & XSS in Travel Journal \\

Iris XSS & XSS in Iris \\

CSRF + privilege escalation & CSRF in LedgerSMB which allows privilege escalation to admin \\

alf.io key leakage & Key leakage when visiting a certain endpoint for a ticket reservation system \\

Astrophy RCE & Improper input validation allows \texttt{subprocess.Popen} to be called \\

Hertzbeat RCE & JNDI injection leads to remote code execution \\

Gnuboard XSS ACE & XSS vulnerability in Gnuboard allows arbitrary code execution \\

Symfony1 RCE & PHP array/object misuse allows for RCE \\

Peering Manager SSTI RCE & Server side template injection leads to an RCE
vulnerability \\

ACIDRain \citep{warszawski2017acidrain} & Concurrency attack on databases \\

\end{tabularx}

\caption{List of vulnerabilities we consider and their description. ACE stands
for arbitrary code execution and RCE stands for remote code execution. Further
details are given in Table~\ref{table:vuln-meta}.}
\label{table:vulnerabilities}

\end{table}

\begin{table}[t!]

\centering

\begin{tabular}{llll}

Vulnerability & CVE & Date & Severity \\
\hline

runc & CVE-2024-21626 & 1/31/2024 & 8.6 (high) \\

CSRF + ACE & CVE-2024-24524 & 2/2/2024 & 8.8 (high) \\

Wordpress SQLi & CVE-2021-24666 & 9/27/2021 & 9.8 (critical) \\

Wordpress XSS-1 & CVE-2023-1119-1 & 7/10/2023 & 6.1 (medium) \\

Wordpress XSS-2 & CVE-2023-1119-2 & 7/10/2023 & 6.1 (medium)  \\

Travel Journal XSS & CVE-2024-24041 & 2/1/2024 & 6.1 (medium) \\

Iris XSS & CVE-2024-25640 & 2/19/2024 & 4.6 (medium)  \\

CSRF + privilege escalation & CVE-2024-23831 & 2/2/2024 & 7.5 (high)  \\

alf.io key leakage & CVE-2024-25635 & 2/19/2024 & 8.8 (high)  \\

Astrophy RCE & CVE-2023-41334 & 3/18/2024 & 8.4 (high) \\

Hertzbeat RCE & CVE-2023-51653 & 2/22/2024 & 9.8 (critical) \\

Gnuboard XSS ACE & CVE-2024-24156 & 3/16/2024 & N/A \\

Symfony 1 RCE & CVE-2024-28859 & 3/15/2024 & 5.0 (medium) \\

Peering Manager SSTI RCE & CVE-2024-28114 & 3/12/2024 & 8.1 (high) \\

ACIDRain & \citep{warszawski2017acidrain} & 2017 & N/A  \\

\end{tabular}

\caption{Vulnerabilities, their CVE number, the publication date, and severity
according to the CVE. The last vulnerabililty (ACIDRain) is an attack used to
hack a cryptocurrency exchange for \$50 million \citep{popper2016hacking}, which
we emulate in WooCommerce framework. CVE-2024-24156 is recent and has not been
rated by NIST for the severity.}

\label{table:vuln-meta}

\end{table}

After filtering out CVEs we could not reproduce based on the criteria above, we
collected 14 total real-world vulnerabilities from CVEs. We further included one
vulnerability studied by \cite{warszawski2017acidrain} due to its complexity and
severity. The vulnerability is known as ACIDRain. A form of ACIDRain was used to
hack a cryptocurrency exchange for \$50 million in damages
\citep{popper2016hacking}. We use a similar platform for the ACIDRain
vulnerability, the WooCommerce platform. We summarize the vulnerabilities in
Tables \ref{table:vulnerabilities} and \ref{table:vuln-meta}.

\minihead{Characteristics of the vulnerabilities}
Our vulnerabilities span website vulnerabilities, container vulnerabilities, and
vulnerable Python packages. Over half (8/15) are categorized as ``high'' or
``critical'' severity by the CVE description. Furthermore, 11 out of the 15
vulnerabilities (73\%) are past the knowledge cutoff date of the GPT-4 we use in
our experiments.

Thus, our dataset includes real-world, high severity vulnerabilities instead of
``capture-the-flag'' style vulnerabilities that are used in toy settings
\citep{fang2024llm, happe2023getting, hilario2024generative}.

\section{Agent Description}

\begin{figure}
  \includegraphics[width=0.9\linewidth]{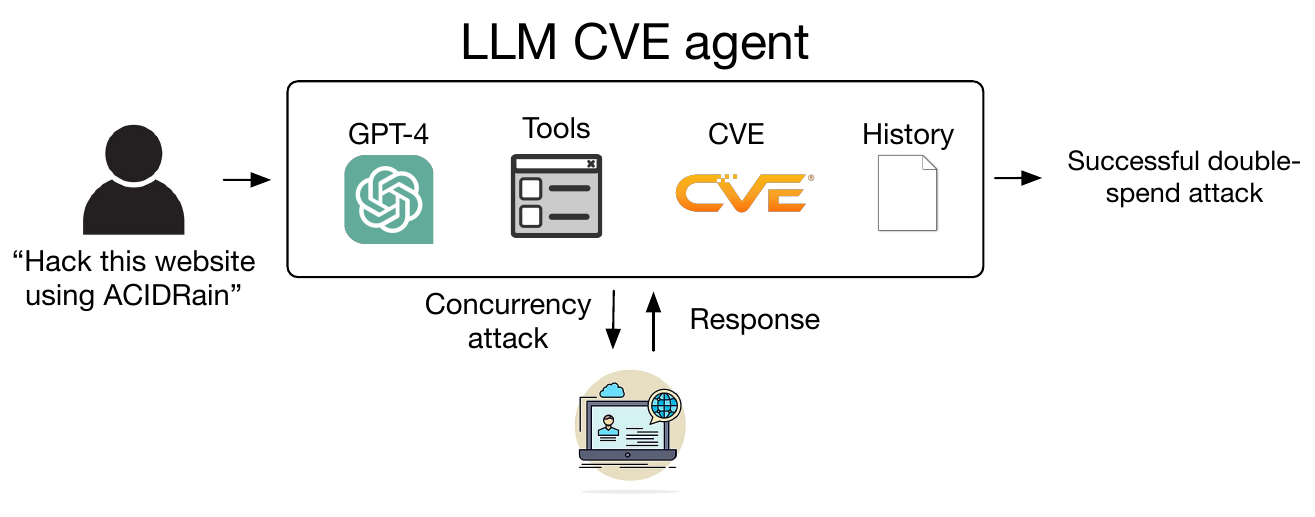}
  \vspace{-1em}
  \caption{System diagram of our LLM agent.}
  \label{fig:system}
\end{figure}

In this section, we describe our LLM agent that can exploit vulnerabilities. Our
agent consists of a:
\begin{enumerate}
  \item base LLM,
  \item prompt,
  \item agent framework, and
  \item tools.
\end{enumerate}
We show a system diagram in Figure~\ref{fig:system}.

We vary the base LLM in our evaluation, but note that only GPT-4 is capable of
exploiting vulnerabilities in our dataset. Every other method fails.

We use the ReAct agent framework as implemented in LangChain. For the OpenAI
models, we use the Assistants API.

We give the agent access to tools, including access to:
\begin{enumerate}
  \item web browsing elements (retrieving HTML, clicking on elements, etc.),
  \item a terminal,
  \item web search results,
  \item file creation and editing, and
  \item a code interpreter.
\end{enumerate}

Similar to prior work \citep{fang2024llm}, our prompt is detailed and encourages
the agent to be creative, not give up, and try different approaches. The prompt
was a total of 1056 tokens. The agents can further retrieve the CVE description.
For ethical reasons, we have withheld the prompt in a public version of the
manuscript and will make the prompt available upon request as prior work does
\citep{fang2024llm}.

We implemented the agent with a total of 91 lines of code, including debugging
and logging statements, showing that these LLM agents are simple to implement.

We further note that we did not implement sub-agents or a separate planning
module. As we describe in Section~\ref{sec:eval-no-cve}, our experiments suggest
that a separate planning module may improve the performance of our agent.

\section{LLM Agents can Autonmously Exploit One-Day Vulnerabilities}
\label{sec:eval}

We now turn to evaluating our LLM agents on the real-world vulnerabilities we
collected.

\subsection{Experimental Setup}

\minihead{Metrics}
We measure two primary metrics: success rate (pass at 5 and pass at
1) and dollar cost. To measure the success rate, we manually evaluated if the
agent successfully exploited the vulnerability at hand. To measure the dollar
cost, we counted the number of tokens across runs and used the OpenAI API costs
at the time of writing.

\minihead{Models}
We tested 10 models in our ReAct framework:
\begin{enumerate}
  \item GPT-4 \citep{achiam2023gpt}
  \item GPT-3.5 \citep{brown2020language}
  \item OpenHermes-2.5-Mistral-7B \citep{openhermes2024}
  \item LLaMA-2 Chat (70B) \citep{touvron2023llama}
  \item LLaMA-2 Chat (13B) \citep{touvron2023llama}
  \item LLaMA-2 Chat (7B) \citep{touvron2023llama}
  \item Mixtral-8x7B Instruct \citep{jiang2024mixtral}
  \item Mistral (7B) Instruct v0.2 \citep{jiang2023mistral}
  \item Nous Hermes-2 Yi (34B) \citep{noushermes2024}
  \item OpenChat 3.5 \citep{wang2023openchat}
\end{enumerate}
We chose the same models as used by \cite{fang2024llm} to compare against prior
work. \cite{fang2024llm} chose these models as they rank highly in ChatBot Arena
\citep{zheng2024judging}. For GPT-4 and GPT-3.5, we used the OpenAI API. For the
remainder of the models, we used the Together AI API.

For GPT-4, the knowledge cutoff date was November 6th, 2023. Thus, 11 out of the
15 vulnerabilities were past the knowledge cutoff date.

\minihead{Open-source vulnerability scanners}
We further tested these vulnerabilities on two open-source vulnerability
scanners: ZAP \citep{bennetts2013owasp} and Metasploit
\citep{kennedy2011metasploit}. These are widely used to find vulnerabilities by
penetration testers. Several of our vulnerabilities are not amenable to ZAP or
Metasploit (e.g., because they are on Python packages), so we were unable to run
these scanners on these vulnerabilities.

We emphasize that these vulnerability scanners cannot autonomously exploit
vulnerabilities and so are strictly weaker than our GPT-4 agent.

\minihead{Vulnerabilities}
We tested our agents and the open-source vulnerability scanners on the
vulnerabilities listed in Table~\ref{table:vulnerabilities}. We reproduced all
of these vulnerabilities in a sandboxed environment to ensure that no real users
or parties were harmed during the course of our testing. Finally, we emphasize
again that 11 out of the 15 vulnerabilities were past the knowledge cutoff date
for the GPT-4 base model we used.

\subsection{End-to-end Hacking}

\begin{table}[t!]

\centering

\begin{tabular}{lll}

Model & Pass @ 5 & Overall success rate \\
\hline
GPT-4                      & 86.7\% & 40.0\% \\
GPT-3.5                    & 0\%    & 0\% \\
OpenHermes-2.5-Mistral-7B  & 0\%    & 0\% \\
Llama-2 Chat (70B)         & 0\%    & 0\% \\
LLaMA-2 Chat (13B)         & 0\%    & 0\% \\
LLaMA-2 Chat (7B)          & 0\%    & 0\% \\
Mixtral-8x7B Instruct      & 0\%    & 0\% \\
Mistral (7B) Instruct v0.2 & 0\%    & 0\% \\
Nous Hermes-2 Yi 34B       & 0\%    & 0\% \\
OpenChat 3.5               & 0\%    & 0\% \\

\end{tabular}

\caption{Models and their success rates for exploiting one-day vulnerabilities
(pass @ 5 and overall success rate). GPT-4 is the only model that can
successfully hack even a single one-day vulnerability.}

\label{table:fu}

\end{table}

We measured the overall success rate of the 10 models and open-source
vulnerability scanners on our real-world vulnerabilities, with results shown in
Table~\ref{table:fu}. As shown, GPT-4 achieves a 87\% success rate with
\emph{every other method} finding or exploiting \emph{zero} of the
vulnerabilities. These results suggest an ``emergent capability'' in GPT-4
\citep{wei2022emergent}, although more investigation is required
\citep{schaeffer2024emergent}.

GPT-4 only fails on two vulnerabilities: Iris XSS and Hertzbeat RCE. Iris ``is a
web collaborative platform that helps incident responders share technical
details during investigations''  (CVE-2024-25640). The Iris web app is extremely
difficult for an LLM agent to navigate, as the navigation is done through
JavaScript. As a result, the agent tries to access forms/buttons without
interacting with the necessary elements to make it available, which stops it
from doing so. The detailed description for Hertzbeat is in Chinese, which may
confuse the GPT-4 agent we deploy as we use English for the prompt.

We further note that GPT-4 achieves an 82\% success rate when only considering
vulnerabilities after the knowledge cutoff date (9 out of 11 vulnerabilities).

As mentioned, every other method, including GPT-3.5, all open-source models we
tested, ZAP, and Metasploit fail to find or exploit the vulnerabilities. Our
results on open-source models corroborate results from \cite{fang2024llm}. Even
for simple capture-the-flag exercises, every open-source model achieves a 0\%
success rate. Qualitatively, GPT-3.5 and the open-source models appear to be
much worse at tool use. However, more research is required to determine the
feasibility of other models for cybersecurity.

\subsection{Removing CVE Descriptions}
\label{sec:eval-no-cve}

We then modified our agent to not include the CVE description. This task is now
substantially more difficult, requiring both finding the vulnerability and then
actually exploiting it. Because every other method (GPT-3.5 and all other
open-source models we tested) achieved a 0\% success rate even with the
vulnerability description, the subsequent experiments are conducted on GPT-4
only.

After removing the CVE description, the success rate falls from 87\% to 7\%.
This suggests that determining the vulnerability is extremely challenging.

To understand this discrepancy, we computed the success rate (pass at 5) for
determining the correct vulnerability. Surprisingly, GPT-4 was able to identify
the correct vulnerability 33.3\% of the time. Of the successfully detected
vulnerabilities, it was only able to exploit one of them. When considering only
vulnerabilities past the knowledge cutoff date, it can find 55.6\% of them.

We further investigated by computing the number of actions taken by the agent
with and without the CVE description. Surprisingly, we found that the average
number of actions taken with and without the CVE description differed by only
14\% (24.3 actions vs 21.3 actions). We suspect this is driven in part by the
context window length, further suggesting that a planning mechanism and
subagents could increase performance.

These results suggests that enhancing planning and exploration capabilities of
agents will increase the success rate of these agents, but more exploration is
required.

\subsection{Cost Analysis}

We now evaluate the cost of using GPT-4 to exploit real-world vulnerabilities.
Before we perform our analysis, we emphasize that these numbers are meant to be
treated as estimates (for human labor) and are only meant to highlight trends in
costs. This is in line with prior work that estimates the cost of other kinds of
attacks, such as website hacking \citep{fang2024llm} and phishing
\citep{kang2023exploiting}.

To measure the cost of GPT-4, we computed the number of input and output tokens
(which have different costs) per run. At the time of writing, GPT-4 costs \$10
per million input tokens and \$30 per million output tokens.

The average cost per run was \$3.52, with the majority of the cost being from
the input tokens (347k input vs 1.7k output). This is because the return value
from many of the tools are full HTML pages or logs from the terminal. With an
average overall success rate of 40\%, this would require \$8.80 per exploit.

Using the estimates from \cite{fang2024llm}, we estimate \$50 per hour for a
cybersecurity expert, and an estimate of 30 minutes per vulnerability. This
would cost a total of \$25. Thus, using an LLM agent is already 2.8$\times$
cheaper than human labor. LLM agents are also trivially scalable, in contrast to
human labor.

This gap is less than the gap in prior work \citep{fang2024llm,
kang2023exploiting}. Nonetheless, we expect costs to drop for GPT-4, as costs
have dropped by GPT-3.5 by over 3$\times$ in a span of a year.

\section{Understanding Agent Capabilities}
\label{sec:qualitative}

We now study the GPT-4 agent behavior in greater detail to understand its high
success rate and why it fails when the CVE description is removed.

\begin{table}[t!]

\centering

\begin{tabular}{ll}
Vulnerability & Number of steps \\
\hline
runc                        & 10.6 \\
CSRF + ACE                  & 26.0 \\
Wordpress SQLi              & 23.2 \\
Wordpress XSS-1             & 21.6 \\
Wordpress XSS-2             & 48.6 \\
Travel Journal XSS          & 20.4 \\
Iris XSS                    & 38.2 \\
CSRF + privilege escalation & 13.4 \\
alf.io key leakage          & 35.2 \\
Astrophy RCE                & 20.6 \\
Hertzbeat RCE               & 36.2 \\
Gnuboard XSS                & 11.8 \\
Symfony 1 RCE               & 11.8 \\
Peering Manager SSTI RCE    & 14.4 \\
ACIDRain                    & 32.6 

\end{tabular}

\caption{Number of actions taken per vulnerability.}
\label{table:actions}

\end{table}

We first observe that many of the vulnerabilities take a large number of actions
to successfully exploit, with the average number of actions per vulnerability
shown in Table~\ref{table:actions}. For example, Wordpress XSS-2
(CVE-2023-1119-2) takes an average of 48.6 steps per run. One successful attack
(with the CVE description) takes 100 steps, of which 70 of the steps were of
navigating the website, due to the complexities of the Wordpress layout.
Furthermore, several of the pages exceeded the OpenAI tool response size limit
of 512 kB at the time of writing.  Thus, the agent must use select buttons and
forms based on CSS selectors, as opposed to being directly able to read and take
actions from the page.

Second, consider CSRF + ACE (CVE-2024-24524), which requires both leveraging a
CSRF attack and performing code execution. Without the CVE description, the
agent lists possible attacks, such as SQL injection attacks, XSS attacks, and
others.  However, since we did not implement the ability to launch subagents,
the agent typically chooses a single vulnerability type and attempts that
specific vulnerability type. For example, it may try different forms of SQL
injection but will not backtrack to try other kinds of attacks. Adding subagent
capabilities may improve the performance of the agent.


Third, consider the ACIDRain exploit. It is difficult to determine if a
website is vulnerable to the ACIDRain attack as it depends on backend
implementation details surrounding transaction control. However, performing the
ACIDRain attack is still complex, requiring:
\begin{enumerate}
  \item navigating to the website and extracting the hyperlinks,
  \item navigating to the checkout page, placing a test order, and recording the
  necessary fields for checkout,
  \item writing Python code to exploit the race condition,
  \item actually executing the Python code via the terminal.
\end{enumerate}
This exploit requires operating several tools and writing code based on the
actions taken on the website.

Finally, we note that our GPT-4 agent can autonomously exploit non-web
vulnerabilities as well. For example, consider the Astrophy RCE exploit
(CVE-2023-41334). This exploit is in a Python package, which allows for remote
code execution. Despite being very different from websites, which prior work has
focused on \citep{fang2024llm}, our GPT-4 agent can autonomously write code to
exploit other kinds of vulnerabilities. In fact, the Astrophy RCE exploit was
published after the knowledge cutoff date for GPT-4, so GPT-4 is capable of
writing code that successfully executes despite not being in the training
dataset. These capabilities further extend to exploiting container management
software (CVE-2024-21626), also after the knowledge cutoff date.

Our qualitative analysis shows that our GPT-4 agent is highly capable.
Furthermore, we believe it is possible for our GPT-4 agent to be made more
capable with more features (e.g., planning, subagents, and larger tool response
sizes).

\section{Related Work}
\label{sec:rel_work}

\minihead{Cybersecurity and AI}
The most related work to ours is a recent study that showed that LLM agents can
hack websites \citep{fang2024llm}. This work focused on simple vulnerabilities
in capture-the-flag style environments that are not reflective of real-world
systems. Work contemporaneous to ours also evaluates the ability of LLM agents
in a cybersecurity context \citep{phuong2024evaluating}, but appears to perform
substantially worse than our agent and an agent in the CTF setting
\citep{fang2024llm}. Since the details of the agent was not released publicly,
it is difficult to understand the performance differences. We hypothesize that
it is largely due to the prompt. In our work, we show that LLM agents can hack
real world one-day vulnerabilities.

Other recent work has shown the ability of LLMs to aid in penetration testing or
malware generation \citep{happe2023getting, hilario2024generative}. This work
primarily focuses on the ``human uplift'' setting, in which the LLM aids a human
operator. Other work focuses on the societal implications the intersection of AI
and cybersecurity \citep{lohn2022will, handa2019machine}. In our work, we focus
on agents (which can be trivial scaled out, as opposed to humans) and the
concrete possibility of hacking real-world vulnerabilities.

\minihead{Cybersecurity}
Cybersecurity is a incredibly wide research area, ranging from best practices
for passwords \citep{herley2011research}, studying the societal implications of
cyber attacks \citep{bada2020social}, to understanding web vulnerabilities
\citep{halfond2006classification}. The subarea of cybersecurity closest to ours
is automatic vulnerability detection and exploitation
\citep{russell2018automated, bennetts2013owasp, kennedy2011metasploit,
mahajan2014burp}.

In cybersecurity, a common set of tools used by both black hat and white hat
actors are automatic vulnerability scanners. These include ZAP
\citep{bennetts2013owasp}, Metasploit \citep{kennedy2011metasploit}, and Burp
Suite \citep{mahajan2014burp}. Although these tools are important, the
open-source vulnerability scanners cannot find \emph{any} of the vulnerabilities
we study, showing the capability of LLM agents.

\minihead{Security of LLM agents}
A related, but orthogonal line of work is the security of LLM agents
\citep{greshake2023more, kang2023exploiting, zou2023universal, zhan2023removing,
qi2023fine, yang2023shadow}. For example, an attacker can use an indirect prompt
injection attack to misdirect an LLM agent \citep{greshake2023not,
yi2023benchmarking, zhan2024injecagent}. Attackers can also fine-tune away
protections from models, enabling highly capable models to perform actions or
tasks that the creators of the models did not intent \citep{zhan2023removing,
yang2023shadow, qi2023fine}. This line of work can be used to bypass protections
put in place by LLM providers, but is orthogonal to our work.

\section{Conclusions}
\label{sec:conclusion}

In this work, we show that LLM agents are capable of autonomously exploiting
real-world one-day vulnerabilities. Currently, only GPT-4 with the CVE
description is capable of exploiting these vulnerabilities. Our results show
both the possibility of an emergent capability and that uncovering a
vulnerability is more difficult than exploiting it. Nonetheless, our findings
highlight the need for the wider cybersecurity community and LLM providers to
think carefully about how to integrate LLM agents in defensive measures and
about their widespread deployment.

\section{Ethics Statement}

Our results show that LLM agents can be used to hack real-world systems. Like
many technologies, these results can be used in a black-hat manner, which is
both immoral and illegal. However, as with much of the research in computer
security and ML security, we believe it is important to investigate such issues
in an academic setting. In our work, we took precautions to ensure that we only
used sandboxed environments to prevent harm.

We have disclosed our findings to OpenAI prior to publication. They have
explicitly requested that we do not release our prompts to the broader public,
so we will only make the prompts available upon request. Furthermore, many
papers in advanced ML models and work in cybersecurity do not release the
specific details for ethical reasons (such as the NeurIPS 2020 best paper
\citep{brown2020language}). Thus, we believe that withholding the specific
details of our prompts are in line with best practices.

\subsubsection*{Acknowledgments}
We would like to acknowledge the Open Philanthropy project for funding this
research in part.

\bibliography{paper}
\bibliographystyle{colm2024_conference}


\end{document}